\documentclass[aps,pra,twocolumn]{revtex4}

\usepackage{epsfig}
\usepackage{amsfonts}
\usepackage{amsmath}
\usepackage{mathbbol}

\newtheorem{theorem}{Theorem}
         
\newtheorem{lemma}{Lemma}          

\newtheorem{definition}{Definition}

\usepackage{hyperref}

\begin{document}

\title{Communication complexity and the reality of the wave-function}

\author{Alberto Montina}

\affiliation{Facolt\`a di Informatica, Universit\`a della Svizzera Italiana,
Via G. Buffi 13, 
6900 Lugano, Switzerland}

\begin{abstract}
In this review, we discuss a relation between  quantum communication
complexity and a long-standing debate in quantum 
foundation concerning the interpretation of the quantum state. Is 
the quantum state a physical element of reality as originally
interpreted by Schr\"odinger? Or is it an abstract mathematical
object containing statistical information about the outcome of
measurements as interpreted by Born? Although these questions
sound philosophical and pointless,
they can be made precise in the framework of what we call classical
theories of quantum processes, which are a reword of quantum phenomena 
in the language of classical probability theory. In 2012, Pusey, Barrett 
and Rudolph (PBR) proved, under an assumption of preparation
independence, a theorem supporting the original interpretation
of Schr\"odinger in the classical framework.
The PBR theorem has attracted considerable interest 
revitalizing the debate and motivating other proofs with alternative
hypotheses. Recently, we showed that these questions are related to a
practical problem in quantum communication complexity, namely,
quantifying the minimal
amount of classical communication required in the classical simulation
of a two-party quantum communication process. In particular, we
argued that the statement of the PBR theorem can be proved if
the classical communication cost of simulating  the communication
of $n$ qubits grows more than exponentially in $n$. Our argument is
based on an assumption that we call probability equipartition property. 
This property is somehow weaker than the preparation
independence property used in the PBR theorem, as the former can
be justified by the latter and the asymptotic equipartition
property of independent stochastic sources. The probability
equipartition property is a general and natural hypothesis that
can be assumed even if the preparation independence hypothesis
is dropped. In this review, we further develop our argument into
the form of a theorem.

\end{abstract}

\maketitle

\section{Introduction}	

One of the main objectives of quantum information theory is to understand
when quantum devices outperform their classical counterpart in terms
of computational resources and amount of communication. The goal can be 
achieved by finding the most efficient model that classically simulates 
the quantum device. Besides the practical interest, the study of
optimal classical simulators can also have
important implications in the context of a long-standing quantum foundational 
debate concerning the interpretation of the quantum state. Is the quantum
state an element of reality, as initially interpreted by Schr\"odinger,
or is it a mere abstract mathematical object of the theory?
As we will discuss in this review, some results 
in quantum communication complexity support a realistic interpretation
of the quantum state in the framework of what we call classical theories
of quantum processes. Let us first discuss the foundational and practical 
motivations that are at the basis of this debate.

Quantum theory provides a consistent 
framework for computing the probabilities of the outcomes of measurements
given some previous knowledge, which is mathematically represented by the 
quantum state. Quantum theory has successfully been applied in fields like
atom physics, 
particle physics, condensed matter and cosmology. In spite of its success, quantum 
theory still suffers from an interpretational issue known as the measurement 
problem. Whenever a particle like an electron is spatially delocalized, 
the formalism does not provide any description of its actual position. 
Although, this feature could be fine in the microscopic world, it
becomes problematic when it is extrapolated to the macroscopic
domain of the every-day experience, as illustrated by Schr\"odinger cat's 
paradox. In the standard interpretation,
the problem is generally fixed by marking a boundary between the fuzzy 
microscopic quantum world and the macroscopic well-defined world made of 
objective observations. 
Besides other interpretations such as the consistent-histories and many-worlds
interpretations,~\cite{c_hist,mwi}
a possible alternative solution of the issue is to fit quantum theory into the
framework of classical probability theory, that is, to remove the
boundary by phagocytizing the quantum domain into the classical one.
In this framework, the state of a system would be described
by a set of classical variables evolving according to some deterministic
or stochastic law. These variables should account for the definiteness
of the macroscopic reality by containing the actual value of what we 
can observe non-invasively, such as the position of a pen on a desk.

As quantum systems can be simulated through classical resources
this reduction to a classical framework is in principle possible.
The simplest way to realize it is to identify the classical
variables with the quantum state, now regarded as an element of reality,
as initially interpreted by Schr\"odinger. The
wave-function would be as real as the waves on the ocean,
a particle being a spatially localized wave-packet.
However, as the wave-function can spread out, this interpretation 
needs some active mechanism that spontaneously localizes the wave-function.
This is the route to realism taken in a collapse theory {\it a la}
Ghirardi, Rimini and Weber~\cite{ghir}. An alternative approach that does
not need a collapse mechanism is taken in pilot-wave theories, where
the quantum state is supplemented by additional auxiliary 
variables, such as the actual position of the particles.
Both collapse theories and pilot-wave theories have the common
feature of promoting the wave-function to the rank of an ontologically 
objective field. For this reason, they are often called 
$\psi$-ontic in the quantum foundation community. 
Curiously, $\psi$-ontic theories are the only currently available 
classical reformulation of quantum theory.  
Unfortunately, this classical reword of quantum theory
does not provide any practical advantage in terms of 
computation of quantum processes. As in the standard formulation,
the computation of a process in the classical model passes through
the solution of the Schr\"odinger equation. Thus, unless $\psi$-ontic
theories are not exactly equivalent to quantum theory and they can 
predict observations detectably different from the standard formulation,
their content remains merely philosophical.

More interestingly, we could wonder if these theories are the only available 
option. As an evidently necessary condition, the classical variables should
contain at least the values of what can be observed non-invasively.
As the quantum state of a single system cannot be directly measured,
there is no evident reason to take it as part of the classical description.
After all, the quantum state can only be recovered from the statistical
distribution of the outcomes of measurements performed on many identically
prepared systems. Thus, the quantum state looks more similar to a probability
distribution, representing our knowledge of what is the actual state
of affairs of a system. Bearing this in mind, in a more general classical
formulation of quantum theory, quantum states could be mapped to
overlapping probability distributions over the classical space, so that the actual
values of the classical variables in a single realization would not contain
the full information about the quantum state. In other words, 
a single statistical realization of the classical variables would be 
compatible with many different quantum states. These hypothetical theories
are called $\psi$-epistemic, since the quantum state is not part of the
ontological description, but it merely represents our statistical knowledge
about the classical variables. The question whether this statistical representation 
is actually possible has attracted growing interest in the recent 
years.~\cite{hardy0,montina,spekkens,montina2,harri,montina3,montina4,bartlett,
montina5,pbr,lewis,colbeck,maxi,hardy} One possible advantage of 
$\psi$-epistemic theories is the fact that the statistical role of the quantum 
state makes them potentially less exposed to the principle of Occam's razor than 
$\psi$-ontic theories. For example, the information required to describe the classical
state of a single system can turn out to be finite on average, whereas the
classical information required to define exactly a quantum state is infinite.
Thus, $\psi$-epistemic theories could be supported by the law of parsimony, as 
suggested in Ref.~\cite{montina} and, more recently, in 
Refs.~\cite{montina4,montina5}. The relevant point is to understand if
$\psi$-epistemic theories exist and if they provide some descriptional advantage 
over their $\psi$-ontic counterpart. We will see that these questions could have a 
negative answer and $\psi$-epistemic theories could collapse
to $\psi$-ontic theories in the asymptotic limit of infinite qubits.

In 2012, Pusey, Barrett and Rudolph~\cite{pbr} (PBR) provided the first proof, under
a hypothesis of preparation independence, that $\psi$-epistemic theories
are incompatible with the predictions of quantum theory. These findings
fed considerable interest and motivated other proofs using alternative hypotheses, 
like in Refs.~\cite{hardy0,colbeck}. Subsequently,
Lewis, Jennings, Barrett and Rudolph reported a counterexample showing that 
the PBR theorem can be evaded once the preparation independence hypothesis
is dropped.~\cite{lewis} However, their findings do not solve definitely our questions.
Indeed, although the reported model is formally $\psi$-epistemic, it has
still some unwanted properties that make it not completely $\psi$-epistemic,
according to the definition given in Ref.~\cite{montina6} and recalled 
later in this review. For example, it occurs that some statistical realizations
of the classical state can still contain the full information about the quantum state.
This implies that the quantum state can be inferred from the classical
state with a finite probability of success. Furthermore, the model collapses
to a $\psi$-ontic model in the limit of infinite qubits.

In Ref.~\cite{montina6}, we showed that the question about the existence of
completely $\psi$-epistemic theories is equivalent to
the quantum communication complexity problem of quantifying the minimal
amount of classical communication required to simulate a two-party quantum 
communication process. More precisely, we showed that a completely $\psi$-epistemic 
theory exists if and only if the communication of qubits can be simulated by a 
classical protocol employing a finite amount of classical communication 
(hereafter, more concisely, {\it finite communication protocol} or {\it FC protocol}).
As the communication of a single qubit can be classically simulated by a 
FC~protocol,~\cite{cerf,toner} completely $\psi$-epistemic models for single qubits exist.
At the present, both FC protocols and 
completely $\psi$-epistemic models are known only for single qubits, as the two problems 
of extending the communication protocols and the $\psi$-epistemic models to more general 
cases are equivalent. 

Provided that this extension is actually possible, we can still wonder how much
the $\psi$-epistemic theory differs from a $\psi$-ontic theory. Indeed, it could
occur that the classical variables turn out to contain the information about
the quantum state up to an error that goes to zero as the number of involved qubits
is increased. In this case, the $\psi$-epistemic theory would collapse to a
$\psi$-ontic theory in the asymptotic limit of infinite qubits (as it occurs
with the model in Ref.~\cite{lewis}).
In Refs.~\cite{montina7,montina8}, we argued that this is the case if
the minimal communication cost of a FC protocol grows more than exponentially
with the number of qubits. This statement can be proved under an assumption that 
we call {\it probability equipartition property}. As we will discuss later, this
property is somehow weaker than the preparation independence hypothesis used
in the PBR theorem. Using two mathematical conjectures, 
we also proved that the communication cost grows at least as $n 2^n$~\cite{montina8}.
An exact proof of this lower bound without conjectures would provide a proof of the PBR
theorem by replacing the preparation independence hypothesis with the aforementioned 
equipartition property. We will also see that this lower bound implies that a 
$\psi$-epistemic theory does not provide a descriptional advantage over $\psi$-ontic
theories, even if the equipartition property is dropped.

This review is organized as follows. In section~\ref{sec_class_theory}, we introduce
the framework of a classical theory of quantum processes and provide a mathematical 
definition of $\psi$-ontic, $\psi$-epistemic and completely $\psi$-epistemic theories.
Section~\ref{section_3} is devoted to classical protocols simulating a two-party
communication process and to the definition of communication cost of a simulation.
In section~\ref{sim_protocol}
we establish a relationship between completely $\psi$-epistemic theories and
protocols with a finite communication cost.
In section~\ref{main_sec},  we show that a
$\psi$-epistemic theory collapses to a $\psi$-ontic theory in the limit
of infinite qubits, under the assumption that the probability equipartition property holds
and the minimal communication cost of a FC protocol grows more than exponentially
in the number of qubits. Finally, the conclusions are drawn.
This short review is mainly focused on some recent results of the Author. An
extensive review of other results in the field can be found in Ref.~\cite{leifer_0}.

\section{Classical Reformulation of Quantum Theory}
\label{sec_class_theory}

Let us introduce the general framework of a classical theory of quantum systems.
By classical theory, we just mean a classical probability theory of quantum processes.
The theory does not necessarily have a structure resembling classical mechanics.
Determinism is neither required. For our purposes, it is sufficient to consider
the simple scenario of state preparation and subsequent measurement.

In the classical theory, a system is described by a set of variables, which we
denote by $x$. As these variables are meant to be ontologically objective, let
us call their actual value the ontic state of the system. 
When the system is prepared in some quantum state $|\psi\rangle$, the preparation
procedure modifies the variable $x$ through some process that sets its
value according to a probability distribution $\rho(x|\psi)$ depending on the procedure,
that is, on $|\psi\rangle$. To simplify the notation, hereafter the ket $|\psi\rangle$
is concisely denoted by $\psi$. The bra-ket notation will be used only for
scalar products.
More generally, the probability distribution could
depend on additional parameters specifying the preparation context, but this is
irrelevant for our discussion. Thus, there is a mapping
\begin{equation}\label{onto_distr}
\psi\rightarrow \rho(x|\psi)
\end{equation}
that associates each quantum state with a probability distribution on the classical
space. The mapping must be injective.
Note that this mapping could be achieved with a classical space much smaller than the 
Hilbert space. Indeed, the minimal number of classical states required to have an 
injective mapping is finite and equal to the double of the Hilbert space dimension,
whereas the number of quantum states is infinite. However this 
minimal requirement is not sufficient to provide an effective classical simulation of the 
overall process of state preparation and measurement.

In quantum theory, a general measurement is described by a positive-operator valued 
measure (POVM), which is defined by a set of positive semidefinite operators, 
$\{\hat E_1,\hat E_2,\dots\}\equiv{\cal M}$. Each operator $\hat E_i$ labels an 
event of the measurement $\cal M$. In the framework of the classical theory, the 
probability of $\hat E_i$ is conditioned by the ontic state $x$. Thus, each measurement
$\cal M$ is associated with a probability distribution $P(\hat E_i|x,{\cal M})$,
\begin{equation}
\label{onto_meas}
{\cal M}\rightarrow P(\hat E_i|x,{\cal M}).
\end{equation}
Also in this case, the conditional probability can depend on additional
parameters specifying the measurement context,~\cite{leifer_0} but we
can safely ignore them.
Finally, the classical theory is equivalent to quantum theory if the probability
of having $\hat E_i$ given the preparation $\psi$ is equal to the quantum
probability, that is,
\begin{equation}
\label{Q_constr}
\int dx P(\hat E_i|x,{\cal M}) \rho(x|\psi)=\langle\psi|\hat E_i|\psi\rangle.
\end{equation}
It is worth to underline that the integral symbol stands for integral over
some manifold. Here and hereafter, we could indifferently replace the manifold
with a more abstract measurable space.

\subsection{$\psi$-ontic and $\psi$-epistemic theories}

As said in the introduction, the simplest (and trivial) way to fit quantum theory into the
framework of classical probability theory is to identify the classical state
with the quantum state. In this case, the classical state $x$ is a vector in the
Hilbert space. Thus, the mapping~(\ref{onto_distr}) is
\begin{equation}
\psi\rightarrow \rho(x|\psi)=\delta(x-\psi).
\end{equation}
The conditional probability of getting $\hat E_i$ given $x$ and the measurement $\cal M$
is trivially
\begin{equation}
\label{collapse_theor}
P(\hat E_i|x,{\cal M})=\langle x|\hat E_i|x\rangle.
\end{equation}
Like in a pilot-wave theory, the model can be made deterministic by adding some 
auxiliary classical variables. In this kind of models, the quantum state takes 
part in the classical description and it is regarded as ontologically objective.
For this reason, such theories are called $\psi$-ontic.~\cite{leifer_0}
In a $\psi$-ontic theory, the ontic state $x$ always contains the full information 
about the quantum state. 
\begin{definition}
({\it strong definition}) A theory is $\psi$-ontic in strong sense if the quantum
state can be inferred from the classical state, that is, if $\rho(x|\psi)$
is a delta distribution in $\psi$ for every $x$.
\end{definition}
There is another definition that is somehow weaker and is widely employed,
such as in the PBR paper.~\cite{pbr}
\begin{definition}
\label{weak_def}
({\it weak definition}) A theory is $\psi$-ontic in weak sense if
the distributions $\rho(x|\psi_1)$ and $\rho(x|\psi_2)$ are not overlapping
for every $\psi_1\ne\psi_2$.
\end{definition}
The difference between these two definitions could look marginal, but it is
not. According to the second definition, a theory is $\psi$-ontic if it is 
possible to infer one of two given quantum states, once the classical
state is known. However, this does not imply that a quantum state can
be inferred from the classical state if we have no {\it a priori}
information about the quantum state. Indeed, the model for single qubits
in Ref.~\cite{montina4} is $\psi$-ontic only according to the second definition.
Under the preparation independence hypothesis, the PBR theorem proves that a 
classical theory is $\psi$-ontic in weak sense, but not in strong sense.
As the weak definition is the most popular one, we also employ it in this
review.

Theories that are not $\psi$-ontic are called $\psi$-epistemic. These theories are 
less trivial than their counterpart and are object of this review.
In a $\psi$-epistemic theory, the information about the quantum state is encoded only in 
the statistical behaviour of $x$, that is, in the distribution $\rho(x|\psi)$. 
It is only required that the mapping~(\ref{onto_distr}) is injective. This feature
takes to the possibility of a reduction of information required to specify
the ontic state. For example, whereas the amount of information required to
specify the classical state of a $\psi$-ontic theory is obviously infinite,
this amount could be finite on average in a $\psi$-epistemic theory.

\subsubsection{Completely $\psi$-epistemic theories}

Let us introduce a subclass of $\psi$-epistemic theories that satisfy a very
reasonable condition. These theories are particularly relevant for the present 
discussion and for their role in quantum communication complexity. The condition
that we are going to introduce is a weaker consequence of the following two natural
conditions. First, we assume that the probability density $\rho(x|\psi)$ is bounded 
by some constant for every $x$ and $\psi$. For example, this property is satisfied 
if the distribution is some smooth function. We also assume that the supports of 
$\rho(x|\psi)$ and $\rho(x)=\int d\psi \rho(x|\psi)\rho(\psi)$ have a finite measure 
for every distribution $\rho(\psi)$. In particular, this is true if the space $x$ is
compact. Under these reasonable conditions, the entropy $H(x|\psi)$ of $x$ given 
$\psi$ is finite, as well as the entropy $H(x)$ [Note the abuse of notation.
$H(x|\psi)$ is not a function of $x$ and $\psi$].
The entropy of $x$ (which can be a set of continuous variables)
is not well-defined, as it depends on the measure taken on the classical space.
A measure-independent quantity is 
\begin{equation}
I(x;\psi)\equiv H(x|\psi)-H(x),
\end{equation}
which is also finite for every $\psi$, that is, the quantity
\begin{equation}
\label{finite_cap}
C(\psi\rightarrow x)\equiv \max_{\rho(\psi)} I(x;\psi)
\end{equation}
is finite. In information theory, $I(x;\psi)$ is known as the mutual information
between $x$ and $\psi$. It quantifies the degree of dependence between two
stochastic variables. The quantity $C(\psi\rightarrow x)$ is the capacity
of the communication channel $\psi\rightarrow x$ associated with the
conditional probability $\rho(x|\psi)$,~\cite{cover} $\psi$ and $x$ being the input and
outcome of the channel, respectively. Let us recall that a channel $y\rightarrow x$ 
is a stochastic process from an input variable $y$ to an output variable $x$ 
described by a conditional probability $\rho(x|y)$. In information theory, a 
channel represents a physical device, such as a wire, carrying information from a 
sender to a receiver. The information-theoretic interpretation of the channel
capacity is provided by the noisy-channel coding theorem.~\cite{cover} Roughly
speaking, the capacity of a channel is the rate of information that can
be transmitted through the channel. Now we define a completely $\psi$-epistemic model by
keeping only property~(\ref{finite_cap}).
\begin{definition}
\label{compl_psi_epist}
The classical model defined by the maps~(\ref{onto_distr},\ref{onto_meas})
is completely $\psi$-epistemic if the capacity of the channel $\psi\rightarrow x$ is 
finite.
\end{definition}
This definition has been justified by assuming that $\rho(x|\psi)$ is bounded and the 
space of $x$ is compact, but the defined class is actually broader, as it includes
some models such that $\rho(x|\psi)$ is not bounded and the supports of
$\rho(x|\psi)$ and $\rho(x)$ do not have finite measure. It is not hard to show
that a completely $\psi$-epistemic theory is $\psi$-epistemic. Furthermore, 
it will become clear in section~\ref{main_sec} that the two classes are equivalent
if the assumption of probability equipartition holds.

\subsubsection{Example: Kochen-Specker model}
\label{kochen-specker}
The Kochen-Specker model~\cite{ks} is an example of completely $\psi$-epistemic model 
working for single qubits. The ontic state 
is given by a unit three-dimensional vector, $\vec x$. Let us represent a pure quantum
state through the unit Bloch vector, $\vec v$. Given the quantum state $\vec v$,
the probability distribution of $\vec x$ is
\begin{equation}
\rho(\vec x|\vec v)=\pi^{-1}\vec v\cdot\vec x\theta(\vec v\cdot\vec x),
\end{equation}
where $\theta$ is the Heaviside step function. As shown in Ref.~\cite{montina6},
the capacity of the channel $\vec v\rightarrow\vec x$ is
$
C(\vec v\rightarrow\vec x)=
2-(2\log_e2)^{-1}\simeq 1.28 \text{ bits}.
$
At the present, no other completely $\psi$-epistemic model is known for
higher dimensional quantum systems.

\section{Communication complexity of a Two-Party Communication Process}
\label{section_3}
In the previous section, we have introduced the general structure of a classical model 
that simulates the quantum process of state preparation and subsequent measurement. 
This process can be regarded as the following communication process between two parties.
A sender, say Alice, chooses a quantum state $\psi$ and sends it to another party, 
say Bob, who performs a measurement chosen by him. 
A problem in quantum communication complexity is to quantify the minimal amount of classical
communication required to simulate the two-party quantum process through a classical
protocol. The classical protocol has the same structure of the classical models introduced
in the previous section, once the ontic variable $x$ is identified with the communicated 
variable and some possible additional stochastic variables shared between
the sender and the receiver. A classical protocol is as follows. Alice chooses a 
state $\psi$ and generates a variable 
$k$ with a probability $\rho(k|\chi,\psi)$ depending on $\psi$ and a possible 
random variable, $\chi$, shared with Bob. The variable $\chi$ is generated according to the 
probability distribution $\rho_s(\chi)$. Note the $\chi$ is independent of $\psi$.
It can be regarded as a distributed key that is generated before Alice and Bob choose
the state $\psi$ and the measurement, respectively, and the protocol is initiated. 
Then, Alice communicates the value of $k$ 
to Bob. Finally, Bob chooses a measurement $\cal M$ and 
generates an outcome $\hat E_i$ with a probability 
$P(\hat E_i|k,\chi,{\cal M})$. The protocol simulates exactly the quantum channel if the
probability of $\hat E_i$ given $\psi$ is equal to the quantum probability,
that is, if
\begin{equation}
\sum_k\int d\chi P(\hat E_i|k,\chi,{\cal M}) \rho(k|\chi,\psi)\rho_s(\chi)=\langle\psi|\hat E_i|\psi\rangle.
\end{equation}
As said, the protocol is equivalent to the model introduced in the previous section, 
the variable $x$ corresponding to the pair $(k,\chi)$. The distribution $\rho(x|\psi)$ 
in the mapping~(\ref{onto_distr}) corresponds to the distribution 
$\rho(k,\chi|\psi)\equiv\rho(k|\chi,\psi)\rho_s(\chi)$. 

There are different definitions of communication cost of a classical simulation.
Without loss of generality, we can assume that $k$ is deterministically 
generated from $\psi$ and $\chi$. If this is not the case, we can make the
protocol deterministic by adding auxiliary stochastic variables that we include 
in $\chi$. The variable $k$ can be regarded as a sequence of bits whose
number depends on $\chi$ and $\psi$.
Let $C(\psi,\chi)$ be the number of bits sent by Alice when the state $\psi$
is chosen with the shared noise $\chi$. The worst-case cost is the maximum of
$C(\psi,\chi)$ over every possible value taken by $\chi$ and $\psi$.
As an alternative, denoting by $C(\psi)$ the average of $C(\psi,\chi)$ 
over $\chi$, we can define the cost as the maximum of $C(\psi)$ over $\psi$. 
Denoting by $\bar {\cal C}$ this quantity, we have
\begin{equation}
\bar {\cal C}\equiv \max_{\psi} \int d\chi \rho_s(\chi) C(\psi,\chi).
\end{equation}
There is also an entropic definition.~\cite{montina_p_w} 
For our purposes, the average and entropic cost can be indifferently used.
Here, we will refer to the average cost $\bar {\cal C}$.
\begin{definition}
We define the {\it communication complexity} ${\cal C}_{min}$ of a quantum communication
process as the minimal amount of classical communication required by an exact classical 
simulation of the process.
\end{definition}

\subsection{Parallel simulations}

If a parallel simulation of $N$ quantum processes are performed, it is possible to envisage a 
larger set of communication protocols, where the probability of generating $k$ can depend
on the full set of quantum states, say $\psi_{i=1,2,\dots,N}$, prepared in 
each single process. In other words, the distribution $\rho(k|\chi,\psi)$ becomes
$\rho(k|\chi,\psi_1,\psi_2,\dots,\psi_{N})$. The asymptotic communication cost, 
${\cal C}^{asym}$, is the cost of the parallelized simulation divided by $N$ in the
limit of large $N$.
\begin{definition}
We define the {\it asymptotic communication complexity} ${\cal C}^{asym}_{min}$ of a quantum 
process as the minimum of ${\cal C}^{asym}$ over the class of parallel protocols
that simulate the process.
\end{definition}
Since the set of protocols working for parallel simulations is larger than the set of 
single-shot protocols, it is clear that
\begin{equation}
{\cal C}_{min}^{asym}\le {\cal C}_{min}.
\end{equation}
However, the difference between ${\cal C}_{min}^{asym}$ and ${\cal C}_{min}$ 
is tiny and not bigger than the logarithm of ${\cal C}_{min}^{asym}$.~\cite{montina_p_w}

\section{$\psi$-Epistemic Theories and Communication Complexity}
\label{sim_protocol}

A finite communication protocol (FC protocol) of a quantum process is a protocol
that simulates the process with a finite amount of classical communication. 
Using the data processing inequality and the chain rule for the mutual
information,~\cite{cover} it is possible to show that a FC protocol corresponds
to a completely $\psi$-epistemic classical model. Let us denote by $I(x;y|z)$
the conditional mutual information between $x$ and $y$ given $z$,
which is the mutual information between $x$ and $y$ given $z$
and averaged on $z$.
From the chain rule~\cite{cover}
\begin{equation}
I(k,\chi;\psi)=I(\chi;\psi)+I(k;\psi|\chi)
\end{equation}
and the fact that $\psi$ and $\chi$ are uncorrelated, we have that
\begin{equation}
I(k,\chi;\psi)=I(k;\psi|\chi).
\end{equation}
From the data-processing inequality, we have that $I(k;\psi|\chi)$ is smaller than
or equal to the communication cost $\bar{\cal C}$ for any $\rho(\psi)$, that is,
\begin{equation}
\bar{\cal C}\ge C \left[\psi\rightarrow (k,\chi) \right],
\end{equation}
where $C\left[\psi\rightarrow (k,\chi)\right]$ is the capacity of the channel 
$\psi\rightarrow (k,\chi)$. Thus, if the communication cost is
finite, the protocol corresponds to a completely $\psi$-epistemic model
with $(k,\chi)$ as ontic variable. Thus, we have the following.
\begin{lemma}
A finite-communication protocol is a completely $\psi$-epistemic classical model.
\end{lemma}
In Ref.~\cite{montina6}, we showed that also the opposite is true in some
sense. Namely, we showed that there is procedure turning a completely 
$\psi$-epistemic model into a finite-communication protocol. 

\subsection{FC protocols from completely $\psi$-epistemic models}
\label{procedure}

We now describe the procedure introduced in Ref.~\cite{montina6} for generating a FC 
protocol from a completely $\psi$-epistemic classical model. This procedure is a consequence 
of the reverse Shannon theorem~\cite{rev_shannon} and its one-shot version~\cite{harsha}. 
Given $M$ identical a noisy channels $x\rightarrow y$, defined by the conditional probability 
$\rho(y|x)$ and with capacity $C_{ch}$, the reverse Shannon theorem states that the channels
can be simulated through a noiseless channel with a communication cost equal to $M C_{ch}+o(M)$,
provided that the sender and receiver share some random variable. 
In other words the asymptotic communication cost of a parallel simulation of many copies of a 
channel $x\rightarrow y$ is equal to $C_{ch}$.
A one-shot version of this theorem was recently reported in Ref.~\cite{harsha}. 
The communication cost $\bar{\cal C}$ of simulating a single channel $x\rightarrow y$
satisfies the bounds
\begin{equation}
\label{cost_one-shot}
C_{ch}\le\bar{\cal C}\le C_{ch}+2\log_2(C_{ch}+1)+2\log_2e.
\end{equation}
Thus, the communication cost is $C_{ch}$ plus a possible small additional 
cost that does not grow more than the logarithm of $C_{ch}+1$. 

These results have an immediate application to the problem of deriving FC protocols from 
completely $\psi$-epistemic model. Let $x$ be the classical variable in the completely
$\psi$-epistemic model (see section~\ref{sec_class_theory}). In general, the direct communication 
of this variable can require infinite bits.
A strategy for making the communication finite and as small as possible 
is as follows. Instead of communicating directly the variable $x$ 
Alice can communicate an amount of information that allows Bob to generate $x$ 
according to the probability distribution $\rho(x|\psi)$. By
Eq.~(\ref{cost_one-shot}), the minimal amount of required communication is
essentially equal to the capacity $C_{ch}$ of the channel $\psi\rightarrow x$.
Since $C_{ch}$ is finite in a completely $\psi$-epistemic model, the communication
cost of the simulation protocol is finite.
If many simulations are performed in parallel, the reverse Shannon theorem implies that
there is a classical simulation such that the asymptotic communication cost
is strictly equal to $C_{ch}$.
\begin{theorem}
\label{theor_psi-epist}
There is a procedure that turns a completely $\psi$-epistemic model into a FC
protocol whose communication cost $\bar{\cal C}$ is bounded by the Ineqs.~(\ref{cost_one-shot}),
where $C_{ch}$ is the capacity of the channel $\psi\rightarrow x$.
In the case of a parallel simulation of many instances, there is a protocol whose
asymptotic communication cost is strictly equal to $C_{ch}$.
\end{theorem}

\subsubsection{Communication cost of simulating the communication of a single qubit}
Theorem~\ref{theor_psi-epist} can be immediately applied to the Kochen-Specker model, introduced in
section~\ref{kochen-specker}. We have seen that the capacity of the channel 
$\psi\rightarrow x$ in this model is equal to about $1.28$ bits. Thus, theorem~\ref{theor_psi-epist}
implies that there is a parallel simulation of many instances of a single qubit
communication process such that the asymptotic communication cost is equal to
$1.28$ bits~\cite{montina6}. This value is lower than the upper bound $1.85$ bits proved by Toner
and Bacon~\cite{toner} in the case of parallel simulations.

\section{$\psi$-Epistemic Theories in the Limit of Infinite Qubits}
\label{main_sec}

In the previous section we have seen that completely $\psi$-epistemic theories and 
finite-communication protocols are two sides of the same coin. To find
a completely $\psi$-epistemic theory means to find a FC protocol and viceversa.
Indeed, note that both completely $\psi$-epistemic theories and FC protocols are not 
yet known, apart from the case of single qubits. Suppose that a completely 
$\psi$-epistemic theory actually exists for any finite number of qubits, 
it could turn out that the difference from a $\psi$-ontic theory is
actually small and the overlap between $\rho(x|\psi)$ and
$\rho(x|\psi')$ could go to zero in the limit of infinite qubits for
every pair $(\psi,\psi')$.
In this case, the $\psi$-epistemic theory
would collapse to a $\psi$-ontic theory in this limit. In this section,
we will prove that this is the case if the communication complexity of
a quantum communication process grows more than exponentially in the
number of communicated qubits, provided that a suitable reasonable equipartition
property is satisfied. This property will be discussed later in the section.
In the proof, we employ the weak definition~\ref{weak_def} of a $\psi$-ontic
theory, which is also  used in the PBR paper.~\cite{pbr}

Let ${\cal C}(n)$ be the asymptotic communication complexity of a quantum communication
process where the sender Alice can prepare any quantum state of $n$ qubits and the receiver
Bob can perform any measurement. We could also 
consider the one-shot communication complexity, as it differs by a small amount that
is irrelevant for the following discussion. As defined in section~\ref{sec_class_theory},
let $\rho(x|\psi)$ be the conditional probability of a generic classical model 
of $n$ qubits. Theorem~\ref{theor_psi-epist} implies the inequality
\begin{equation}
\label{bound_mutual}
{\cal C}(n)\le C(\psi\rightarrow x),
\end{equation}
$C(\psi\rightarrow x)$ being the capacity of the process $\psi\rightarrow x$.
By definition of channel capacity, we have that
\begin{equation}
\label{ch_capa}
C(\psi\rightarrow x)=
\max_{\rho(\psi) }\int dx\int d\psi \rho(x,\psi)\log_2
\frac{\rho(x|\psi)}{\rho(x)}.
\end{equation}

In a general $\psi$-epistemic theory, the conditional probability
$\rho(x|\psi)$ can be a mixture of a broad smooth function and very 
narrow functions. If the position of these narrow peaks does not
depend on the quantum state, we can remove them with a change of
the measure on the classical space manifold. In the antipodal
case that their position is a bijective function of the quantum state, 
it turns out that the quantum state 
can be inferred with a very small error and a finite probability of 
success. In this case, the theory is somehow partially $\psi$-ontic.
For example, the model in Ref.~\cite{lewis}, which is 
formally $\psi$-epistemic (but not completely $\psi$-epistemic), displays 
this feature. 
Although this kind of theory is, technically speaking, $\psi$-epistemic, 
it looks very artificial and unpalatable, especially if we are not
interested to find a mere classical simulation of quantum processes
with possible practical interest in quantum information theory, but we pretend
to find a classical theory picturing what actually occurs in the
backstage of quantum phenomena. A classical theory of a complex system
made of a high number of qubits should satisfy a natural requirement that
we call probability equipartition property (or, more exotically,
ontological equipartition property). As we will see, this property is 
related to the asymptotic equipartition property known in information 
theory~\cite{cover} and it is stated as follows.
\begin{definition}
\label{onto_equip}
A classical theory satisfies the probability (or ontological) 
equipartition property
if, given $\psi$, there is a typical set of classical states with probability 
close to one such that the probability distribution $\rho(x|\psi)$
is approximately a constant independent of $\psi$, in the limit of a 
high number of qubits.
\end{definition}
For our purposes, the probability equipartition property can be
satisfied very roughly and some deviation from the uniformity
can be acceptable. It is sufficient that $\rho(x|\psi)$ has
the same order of magnitude on the typical set of classical 
states. This property is introduced to discard theories displaying
huge narrow fluctuations in the probability distribution.
Clearly, the model in Ref.~\cite{lewis} does not satisfy the
equipartition property, as the corresponding distribution is the
mixture of a broad function and a delta distribution.
Furthermore, we assume that the marginal distribution 
$\rho(x)=\int d\psi \rho(x|\psi)\rho(\psi)$ 
satisfies the uniformity property of $\rho(x|\psi)$ for a
uniform distribution $\rho(\psi)$ of quantum states.
This assumption is very reasonable, as
$\rho(x)$ is the probability distribution of the classical state
provided that nothing is known about the quantum state. Again,
the uniformity can be satisfied roughly.

The ontological equipartition property is somehow weaker than the
preparation independence property used in the PBR theorem. Indeed,
the latter justifies the former. A procedure for preparing a general
quantum state of $n$ qubits is as follows. First, we prepare each
qubit in the same quantum state, then we let them evolve according
to some suitable unitary evolution. The unitary evolution can be
implemented through some quantum circuit. Let us consider the
first stage of this procedure. Under the preparation 
independence hypothesis of the PBR theorem, each qubit is associated
with a classical variable $x_i$ and the collection of these variables
is the overall classical state $x$. Furthermore the preparation 
independence property claims that the variables $x_i$ are independent 
stochastic variables, provided that the qubits are prepared in a factorized
quantum state. Thus, we can conclude that the probability distribution
$\rho_0(x)$ after the first stage of the quantum state preparation
satisfies the uniformity property of definition~\ref{onto_equip}.
Indeed, this is a consequence of the asymptotic equipartition property
of independent stochastic processes.~\cite{cover} Let us consider
the second stage of the quantum state preparation. As a unitary evolution
is a reversible conservative process, we can argue that also the
associated classical process describing the evolution of $x$ is
somehow conservative, in the sense that the volume of sets in
the classical space is somehow conserved during the evolution.
More generally, we can argue that the process conserves the uniformity 
property of the initial distribution $\rho_0(x)$, implying that
the distribution $\rho(x|\psi)$ satisfies the uniformity property 
for every $\psi$. That is, $\rho(x|\psi)$ is approximately equal
to a constant independent of $\psi$ in the typical set. The uniformity 
property is a 
general feature of complex systems with a high number of variables and 
can be reasonably assumed even if the preparation independence property
is dropped.

The probability equipartition property and the uniformity of
$\rho(x)$ for a uniform $\rho(\psi)$ imply that the maximum in 
Eq.~(\ref{ch_capa}) is achieved for a uniform distribution $\rho(\psi)$.
This can be verified by using the Karush-Kuhn-Tucker conditions for
optimality. Let us choose a measure in the space of $\psi$ such that the 
uniform distribution is
\begin{equation}
\label{norm_rho_p}
\rho(\psi)=1.
\end{equation}
At this point, by taking a suitable measure on the manifold of $x$ so that 
the function $\rho(x|\psi)$ is equal to one on its support, it is easy to 
prove that 
\begin{equation}
\label{overl_capa}
\int d\psi'\rho(\psi')\omega(\psi,\psi')\simeq 2^{-C(\psi\rightarrow x)},
\end{equation}
where 
\begin{equation}
\omega(\psi,\psi')\equiv \int dx\rho(x|\psi)\rho(x|\psi')
\end{equation}
is the overlap between $\rho(x|\psi)$ and $\rho(x|\psi')$.
The overlap definition can be recast in the measure-independent form
\begin{equation}
\omega(\psi,\psi')=\int dx \min[\rho(x|\psi),\rho(x|\psi')].
\end{equation}
Note that Eq.~(\ref{overl_capa}) implies that a $\psi$-epistemic theory is equivalent 
to a completely $\psi$-epistemic theory if the probability equipartition
property holds.
Using Ineq.~(\ref{bound_mutual}), we have that
\begin{equation}
\label{lower_bound}
\int d\psi'\rho(\psi')\omega(\psi,\psi')\lesssim 2^{-{\cal C}(n)}.
\end{equation}
Thus, the overlap $\omega(\psi,\psi')$, averaged over $\psi'$, goes
to zero for $n\rightarrow\infty$, regardless of how fast the 
communication complexity ${\cal C}(n)$ grows by increasing the 
number of qubits. In other words, most of the pairs of probability
distributions are almost non-overlapping in high dimension. This
feature is still compatible with $\psi$-epistemic theories.
Indeed, two orthogonal quantum states can be always distinguished
by a measurement, implying that their associated probability 
distributions cannot overlap. This can be easily inferred from
the general structure of a classical theory introduced in 
section~\ref{sec_class_theory}. Now, by the principle of the 
concentration of measure, most of the quantum states $\psi'$ are 
almost orthogonal to $\psi$ in high dimension, implying that the 
overlap $\omega(\psi,\psi')$ is almost zero for most of the pairs.

To prove that a $\psi$-epistemic theory collapses to a
$\psi$-ontic theory, we have to show that the overlap 
$\omega(\psi,\psi')$ goes to zero for $n\rightarrow\infty$
regardless how close the states $\psi$ and $\psi'$ are.
To do this,
we single out pairs of quantum states whose distance
is bounded above by any given constant, that is, whose scalar product
is not smaller than any given constant. Let $S(\theta)$ be a set
of quantum states $\psi'$ satisfying the constraint
$|\langle\psi|\psi'\rangle|^2\ge \cos^2\theta$ for some given
vector $\psi$ and angle $\theta$. This set is a kind of cap
whose angular aperture is $2\theta$. 
The volume, say $\Omega_S$, of the set $S$ is 
\begin{equation}
\label{cap_v}
\Omega_S=2^{(2^n-1)\log_2 sin\theta}
\end{equation}
(Note that the volume of the whole quantum state manifold is equal
to $1$).

Now, we perform the integral in Ineq.~(\ref{lower_bound}) only
on the set $S(\theta)$. Obviously, the inequality is still satisfied,
that is,
\begin{equation}
\label{bound_overl}
\int_{S(\theta)} d\psi'\rho(\psi')\omega(\psi,\psi')\lesssim 2^{-{\cal C}(n)}
\end{equation}
for every $\theta\in[0,\pi/2]$. Dividing both sides by the
integral of $\rho(\psi')$ over the set $S$ and bearing in
mind that $\rho(\psi)=1$ [see Eq.~(\ref{norm_rho_p})],
we have from Eq.~(\ref{cap_v}) and Ineq.~(\ref{bound_overl}) that 
\begin{equation}
\bar\omega(\theta)\lesssim 2^{-{\cal C}(n)+(1-2^n)\log_2 sin\theta},
\end{equation}
where $\bar\omega(\theta)$ is the average value of 
$\omega(\psi,\psi')$ over the set $S(\theta)$ of vectors $\psi'$.

This inequality implies that the $\psi$-epistemic theory collapses
to a $\psi$-ontic theory if the communication complexity grows
faster than $2^n$ (according to the weak definition~\ref{weak_def} of
$\psi$-ontic theory). Indeed, in this case, the
right-hand side of the inequality goes to zero for every $\theta$,
that is,
\begin{equation}
\lim_{n\rightarrow\infty}\bar\omega(\theta)=0,\;\; \forall
\theta\in[0,\pi/2].
\end{equation}
In other words, the overlap $\omega(\psi,\psi')$
converges to zero over $S(\theta)$ for every $\theta\in[0,\pi/2]$
in the limit $n\rightarrow\infty$ (mathematically speaking, the convergence
is uniform). Regardless of how close two quantum
states $\psi$ and $\psi'$ are, their overlap goes to zero in the
limit of infinite qubits.
Summarizing, we have the following.
\begin{theorem}
\label{main_theorem}
If the probability equipartition property is satisfied, then
a $\psi$-epistemic theory collapses to a $\psi$-ontic theory
(in the weak sense of definition~\ref{weak_def}),
provided that 
$\lim_{n\rightarrow\infty} 2^{-n}{\cal C}(n)=\infty$.
\end{theorem}

What is known about ${\cal C}(n)$? Brassard, Cleve and Tapp proved
the lower bound $0.01\times 2^n$ for the communication complexity.~\cite{brassard}
Subsequently, the bound was increased to $0.293\times 2^n$ and,
with a mathematical conjecture, to $2^n$~\cite{montina9}.
Unfortunately, these bounds are at the border of the condition stated in 
theorem~\ref{main_theorem}. Very recently, we showed
that ${\cal C}(n)$ scales at least as $n 2^n$ if two
suitable mathematical conjectures hold.~\cite{montina10} The proof 
and the conjectures are quite technical, thus we will not discuss
them in this review. The details can be found in the cited paper. As 
implied by theorem~\ref{main_theorem}, an exact proof of this lower bound 
would provide a proof of the PBR theorem by replacing the independence
hypothesis with the probability equipartition hypothesis. It is worth
to underline that the lower
bound $n2^n$ has another relevant consequence. It is known that
an approximate classical description of the quantum state requires
an amount of information that grows as $n2^n$ up to some factor
that grows as the logarithm of the inverse of the error.
Thus, if the lower bound $n2^n$ holds, then a $\psi$-epistemic
theory would not provided a significant descriptional advantage over an
error-bounded $\psi$-ontic theory, regardless of the probability
equipartition hypothesis.

Our work has a relation with a recent result by Leifer.~\cite{leifer}
If we assume that ${\cal C}(n)$ grows as $2^n$, then Ineq.~(\ref{lower_bound}) 
is similar to an inequality derived by Leifer. As 
$\lim_{n\rightarrow\infty} 2^{-n}{\cal C}(n)$ is finite, the result of 
Leifer is not sufficient to prove that the overlap $\omega(\psi,\psi')$ goes 
to zero for every pair of quantum states. It is interesting to observe that 
Leifer's result comes from the Frankl-R\"odl theorem~\cite{frankl}, which is
also used to prove the bound ${\cal C}(n)>0.01\times 2^n$.~\cite{brassard}

\section{Conclusion}

In principle, quantum theory can be reformulated in the framework
of classical probability theory. The simplest way to reword quantum
phenomena in a classical language is to employ the quantum state
as part of the classical description, possibly supplemented
by auxiliary variables. This is done in the so-called $\psi$-ontic 
theories. However, these reformulations do not provide any substantial 
new content or improvement, unless they can predict observations 
detectably different from the standard formulation. 

A classical theory of quantum processes becomes interesting if it provides substantial
descriptional advantages. In this review, we have discussed about a 
possible alternative class of theories called $\psi$-epistemic.
In spite of their exotic name, these theories are related to 
certain classical protocols studied in quantum communication 
complexity. This relation was first noted in Ref.~\cite{montina6}.
As the quantum state is not part of the classical description,
$\psi$-epistemic theories can potentially introduce some
simplification in the description of quantum systems. 
However, under the assumption of preparation independence, the
PRB theorem implies that such reformulations are incompatible
with quantum theory. Here, we have shown that it is possible
to reach the same conclusion by replacing the preparation
independence property with the somehow weaker equipartition
property. As necessary requirement of the proof, the minimal 
amount of classical communication ${\cal C}(n)$ required to 
replace the communication of $n$ qubits should increase faster
than $2^n$. Interestingly, some recent results suggest that 
${\cal C}(n)$ increases as $n 2^n$.~\cite{montina8} An exact proof of 
this partial result would provide a strong suggestion that 
$\psi$-epistemic theories are actually incompatible with quantum 
theory. 

We conclude by noting that the model in Ref.~\cite{lewis}
collapses to a $\psi$-ontic theory in the limit of infinite
qubits, even if the model does not satisfy the probability
equipartition property. This leads us to wonder if the
collapse to a $\psi$-ontic theory can be proved without
any assumption.

\section*{Acknowledgments}
The Author acknowledge useful discussions with Matthew Leifer,
Jonathan Barrett and Stefan Wolf. This work is supported by
the COST action on Fundamental Problems in Quantum Physics.

\end{document}